%% file: sigconf.tex
\documentclass[sigconf]{acmart}

\AtBeginDocument{%
  }

\setcopyright{acmcopyright}
\copyrightyear{2018}
\acmYear{2018}
\acmDOI{10.1145/1122445.1122456}

\acmConference[SBES'24]{Brazilian Symposium on Software Engineering}{September 30 -- October 04,
  2024}{Curitiba, PR}

\setcopyright{none} 
\renewcommand\footnotetextcopyrightpermission[1]{} 

\usepackage{url}
\usepackage{graphicx}
\usepackage{graphics}
\usepackage{colortbl}
\usepackage{multirow}
\usepackage{listings}

\newcommand{\lstbg}[3][0pt]{{\fboxsep#1\colorbox{#2}{\strut #3}}}

\lstdefinelanguage{diff}{
  basicstyle=\ttfamily\small,
  morecomment=[f][\lstbg{red!20}]-,
  morecomment=[f][\lstbg{green!20}]+,
  morecomment=[f][\textit]{@@},
  morecomment=[f][\textit]{---},
  morecomment=[f][\textit]{+++},
}

\newcommand{\revision}[1]{{\color{black}#1}}

\begin{document}

\title{Assessing Python Style Guides: An Eye-Tracking Study with Novice Developers}


 \author{Pablo Roberto}
 \orcid{0000-0000-0000-0000}
 \affiliation{%
   \normalsize \institution{Federal University of Campina Grande} \country{Brazil}
 }
 \email{pablo@copin.ufcg.edu.br}
 
 \author{Rohit Gheyi}
 \orcid{0000-0002-5562-4449}
 \affiliation{%
   \normalsize \institution{Federal University of Campina Grande} \country{Brazil}
 }
 \email{rohit@dsc.ufcg.edu.br}

 \author{José Aldo Silva da Costa}
 \orcid{0000-0001-8918-1749}
 \affiliation{%
   \normalsize \institution{State University of Paraíba} 
   \country{Brazil}
 }
 \email{jose.aldo@servidor.uepb.edu.br}
 
  \author{Márcio Ribeiro}
 \orcid{0000-0002-4293-4261}
 \affiliation{%
   \normalsize \institution{Federal University of Alagoas} \country{Brazil}
 }
 \email{marcio@ic.ufal.br}

\begin{abstract}

The incorporation and adaptation of style guides play an essential role in software development, influencing code formatting, naming conventions, and structure to enhance readability and simplify maintenance. However, many of these guides often lack empirical studies to validate their recommendations. Previous studies have examined the impact of code styles on developer performance, concluding that some styles have a negative impact on code readability. However, there is a need for more studies that assess other perspectives and the combination of these perspectives on a common basis through experiments. This study aimed to investigate, through eye-tracking, the impact of guidelines in style guides, with a special focus on the PEP8 guide in Python, recognized for its best practices. We conducted a controlled experiment with 32 Python novices, measuring time, the number of attempts, and visual effort through eye-tracking, using fixation duration, fixation count, and regression count for four PEP8 recommendations. Additionally, we conducted interviews to explore the subjects' difficulties and preferences with the programs. The results highlighted that not following the PEP8 Line Break after an Operator guideline increased the eye regression count by 70\% in the code snippet where the standard should have been applied. Most subjects preferred the version that adhered to the PEP8 guideline, and some found the left-aligned organization of operators easier to understand. 
The other evaluated guidelines revealed other interesting nuances, such as the True Comparison, which negatively impacted eye metrics for the PEP8 standard, although subjects preferred the PEP8 suggestion. We recommend practitioners selecting guidelines supported by experimental evaluations.

\end{abstract}

%
%



\maketitle

\input{source/1-introduction}
\input{source/2-background}
\input{source/3-study-definition}
\input{source/4-methodology}

\input{source/5-results}

\input{source/6-threats}
\input{source/7-related-work}

\input{source/8-conclusions}


\section*{Acknowledgments}
\revision{
We would like to thank the anonymous reviewers for their insightful suggestions.}
This work was partially supported by CNPq and FAPEAL grants.

\bibliographystyle{ACM-Reference-Format}
\bibliography{base}
\end{document}

%% file: source/1-introduction.tex
\section{Introduction}
\label{sec:introduction}

Style guides are essential in software development, guiding code formatting, naming conventions, and source code structure to promote readability and facilitate maintenance. Previous studies have investigated the influence of coding styles on readability, with mixed results, indicating that certain styles may compromise code clarity ~\cite{dos2018impacts}. Major companies, such as Google ~\cite{google-styleguide} and Microsoft ~\cite{microsoft-python-formatting}, emphasize style standardization by incorporating guidelines into their corporate style guides, adopting practices aligned with PEP8 to improve code readability and organization.

However, the diversity of programming languages prevents a universal set of style rules, with each language having its own definitions ~\cite{allamanis2014learning}. PEP8, for example, is a widely accepted style guide for Python that suggests best coding practices ~\cite{dasgupta2017code}. Yet, many guides, including PEP8, lack empirical studies as a foundation ~\cite{santos2021estudo}. Previous studies have examined the impact of coding styles ~\cite{bauer2019indentation} on the performance of novice developers and visual effort, but there is a need for more rigorous assessments that consider the developer's perception and how eye transitions may indicate a greater visual effort with certain style patterns.

Conformity with well-established style practices is crucial to ensuring code quality ~\cite{schankin2018descriptive}. PEP8 provides suggestions for coding styles in Python, one of the most popular programming languages. While the PEP8 guide provides justifications for its recommendations, it is crucial to empirically evaluate these recommendations, particularly from dynamic perspectives that consider human factors. For instance, using eye-tracking methodology can provide valuable insights into assessing visual effort~\cite{da2023seeing,costa2021evaluating,oliveira2020atoms}.

The guidelines outlined in PEP8 not only provide suggestions for improvements but also provide code examples in Python of both incorrect and correct code, as shown in the following listing (Listing ~\ref{lst:wrong} and Listing~\ref{lst:correct}) taken from the PEP8 guide. The code in Listings ~\ref{lst:wrong} and ~\ref{lst:correct} present the guideline regarding the use of line breaks before or after the operator. In the \textit{PEP8 compliant} version (see Listing~\ref{lst:correct}), the line break should occur before the operator, not after, as in the \textit{PEP8 non-compliant} version (see Listing ~\ref{lst:wrong}).

\noindent 
  \begin{minipage}[t]{.23\textwidth} 
    \begin{lstlisting}[caption={\textit{PEP8 non compliant} }, label={lst:wrong}]
income = (gross_wages +
          taxable_interest +
          (dividend - quali) -
          ira_deduction -
          stud_interest)
    \end{lstlisting}
  \end{minipage}\hfill 
  \begin{minipage}[t]{.23\textwidth} 
    \begin{lstlisting}[caption={\textit{PEP8 compliant}}, label={lst:correct}]
income = (gross_wages
          + taxable_interest
          + (dividend - quali)
          - ira_deduction
          - stud_interest)
    \end{lstlisting}
  \end{minipage}

For the guideline presented in Listing ~\ref{lst:wrong}, the recommendation states the following: ``the eye has to do extra work to figure out which items are added and which are subtracted''~\cite{van2001pep}. This justification from the style guide led us to question what extra effort the creators of the guide were referring to. In this sense, it becomes important to empirically evaluate the recommendations from the style guide. In particular, we have to consider eye tracking methodology to assess the eye effort that a particular pattern may generate. 

In this study, we employed eye-tracking metrics and conducted interviews with Python novices (32 undergraduate students) to evaluate the impact of four PEP8 guidelines on visual effort and code readability. Our goal was to explore the nuanced relationship between coding patterns and their readability, while also considering the subjective perceptions of developers regarding style preferences and how visual engagement with code might highlight readability issues. Findings suggest that even minor coding patterns recommended by style guides can affect code comprehension, highlighting the need for a more nuanced analysis that includes both objective metrics and the developers' subjective perception.

We evaluate four guidelines (patterns) from the PEP8 style guide: \textit{Whitespace}, \textit{Line Break Before Operator},\textit{ Multiple Statements on the Same Line}, and \textit{Comparison with True}. These guidelines are assessed using a combination of traditional metrics in code comprehension such as time and correct responses, along with eye-tracking metrics including the number of fixations, fixation duration, eye movement regressions, and gaze transitions. Additionally, we conducted interviews with 32 Python novices to understand their preferences and reasons for a particular coding style and correlate them with the results of the assessed metrics.

The obtained results shed light on the potential challenges posed by certain PEP8 guidelines in the context of Python code readability for novices. The findings suggest that adherence to PEP8 guidelines may not always correlate with improved performance, as evidenced by two out of the four evaluated guidelines showing better developer performance in standardized code without PEP8 guidelines. This raises questions about the effectiveness of certain PEP8 recommendations in enhancing code readability, particularly for novice programmers.

This study makes the following contribution:
\begin{itemize}
    \item An eye tracking controlled experiment with 32 novices in Python to investigate the impact of four guidelines from the PEP8 style guide (Section~\ref{method});
    \item A discussion on the quantitative and qualitative eye-tracking results for the four guidelines from the PEP8 style guide (Section~\ref{results-discussion}).
\end{itemize}

\revision{Additionally, the study suggests the need to revisit and possibly update style guidelines like PEP8 based on empirical data and the practical experiences of developers. By incorporating direct feedback from users and results from studies like this one, guidelines can be refined to better meet the needs of modern developers, balancing code clarity and efficiency with ease of learning and use for new programmers.}

%% file: source/2-background.tex
\section{Code Style and Readability}\label{background}

In Software Engineering, \textit{readability} is related to code clarity, i.e., how easy it is to understand the written expression of the code. Almeida et al. ~\cite{de2003best} assert that readability is crucial for code maintenance; if the source code is written in a complex manner, the process of understanding the code will require more effort from the reader. This can result in difficulties in identifying bugs, implementing new features, and making modifications. On the other hand, more readable code tends to be easier to modify and debug, making it more sustainable in the long term. Meanwhile, Daka et al. ~\cite{daka2015modeling} indicate that the visual appearance of code, or style, is generally designated as its readability. The visual organization of code, including formatting, the use of white spaces, and consistency in naming, can significantly affect how developers interpret and interact with the source code. Similarly, Buse and Weimer ~\cite{buse2010learning} developed a metric for code readability, demonstrating that certain characteristics of the code can significantly influence how it is understood. This metric considers factors such as the complexity of control structures, clarity in variable and function naming, and code conciseness. Thus, code that follows best practices in readability tends to be more understandable, facilitating the development, maintenance, and collaboration process among team members.

Buse and Weimer~\cite{buse2008metric,buse2010learning} and Posnett et al.~\cite{posnett2011simpler} aimed to identify specific source code characteristics that directly impact its readability and comprehensibility. These features were assessed through the subjective perceptions of students and programmers, offering valuable insights into factors contributing to code's legibility. Furthermore, Lawrie et al.~\cite{lawrie2006name,lawrie2007effective} explored code identifiers, examining how different naming styles could affect programmers' ease of understanding. These collective research efforts highlight the importance of deliberate and well-founded coding practices to enhance source code accessibility and maintenance, suggesting that seemingly minor details, like identifier choices, can significantly influence code readability. Our work corroborates findings in this field, focusing on the readability of small code snippets within PEP8 guidelines, with participant subjectivity serving as a key evaluation metric.

%% file: source/3-study-definition.tex
\section{Study Definition}
\label{study-definition}


Following the Goal Question Metric approach~\cite{basili-1994}, we aim to analyze Python programs that are PEP8 compliant versus non-compliant for the purpose of comparing them
with respect to their impact on code comprehension from the point of view of novices in Python programming language in the context of tasks adapted from introductory programming courses.

We address the following Research Questions (RQs). Our null hypothesis for each RQ is that there is no difference between the \textit{PEP8 compliant} and \textit{PEP8 non-compliant} versions concerning the collected metric.

\begin{itemize}
    \item \textbf{RQ$_{1}$: To what extent do the PEP8 guidelines affect the task completion time?} To answer this question, we measured how much time the subject needed to specify the correct output for the task. We also measured how much time the subject spent only in specific areas of the code.
    \item \textbf{RQ$_{2}$: To what extent do the PEP8 guidelines affect the number of answer attempts?} To answer this question, we measured the number of attempts the subject made until answering the task correctly, having in mind that the subject is free to make as many attempts as needed.
    \item \textbf{RQ$_{3}$: To what extent do the PEP8 guidelines affect the fixations duration and count?} To answer this question, we measured the number and duration of each fixation found in the data captured from the novices. In the code comprehension scenario, fixations with high duration have been associated with an increase in the level of attention~\cite{busjahn2015eye}. A large number of fixations has been associated with more time to process and understand code statements~\cite{binkley2013impact}, increased attention to complex code~\cite{CR02roles}, and more visual effort to remember identifier names~\cite{sharafi2012women}.
    \item \textbf{RQ$_{4}$: To what extent do the PEP8 guidelines affect the regressions count?} Just as Rayner~\cite{rayner1998eye} observed in reading, regressions can indicate misunderstanding, a concept applied to programming by Busjahn et al.~\cite{busjahn2015eye}. The study assessed regressions in PEP8 code, measuring backward saccades to quantify readability.
    
\end{itemize}

%% file: source/4-methodology.tex
\section{Methodology}\label{method}

\begin{figure}[]%
    \centering
    \qquad
    {{\includegraphics[width=8.5cm]{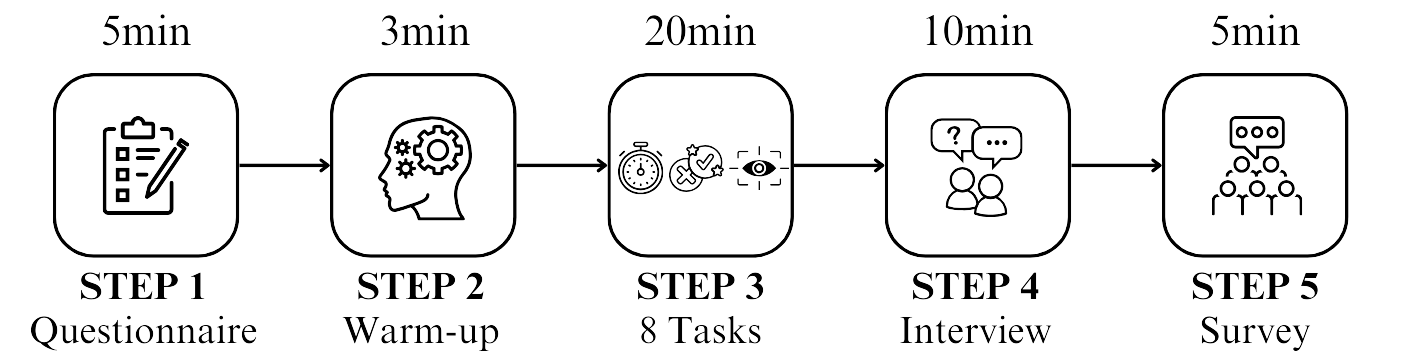} }}%
    \caption{Experiment Steps: Questionnaire, Warm-up, Task, Interview, and Survey.
    }%
    \label{fig: phases}%
\end{figure}

The study was structured in five steps (see Figure~\ref{fig: phases}). Initially, we gave the participants a questionnaire to assess their proficiency in Python and introduced them to relevant code examples. The participants were informed about the best posture for eye-tracking data capturing and reminded of their option to leave the study at any time. Following these preparatory steps, we detailed the eye-tracking camera calibration process to the participants, ensuring they were comfortable and correctly positioned for the procedure. This calibration involved participants following on-screen cues to guarantee accurate eye movement tracking, with recalibrations performed as necessary to ensure data reliability.

In the second step, we simulated the execution of the experiment with a simple warm-up task. While they solved the task, we demonstrated how subjects could specify the output, how the subject could close their eyes for two seconds before and after solving the task, how we signaled the correct and incorrect answers, and how we signaled the time limit. The idea is that the subject can feel comfortable with the experiment setup and the equipment.

In the third step, we conducted the actual experiment with eight programs, half adhering to PEP8 guidelines and the other half functionally equivalent but non-compliant with PEP8 guidelines. To avoid learning effects, we used a Latin Square design~\cite{BO05ST}.

In the fourth step, we concluded the experiment with a semi-structured interview. The goal was to obtain qualitative feedback on how subjects examined the programs and their subjective impressions. We went through each of the eight programs and asked three questions: (1) How difficult was it to find the output: very easy, easy, neutral, difficult, or very difficult? (2) Why this perception? (3) How did you find the output?

Finally, we applied a survey in which we presented code excerpts highlighting the use of the PEP8 guideline or not, and we asked the subject's preference, the motivation for the preference, or if they were indifferent to the use of any of those compared excerpts. We were careful with environmental aspects to reduce noise in the data. For example, we did not use a swivel chair because, in previous pilot studies, subjects tended to move, reducing the accuracy of the eye-tracking equipment. Despite the measures we took, obtaining perfect data is virtually impossible, given the limitations of the camera. Therefore, the collected data were processed, analyzed, and interpreted, correcting the data whenever necessary.

\subsection{Subjects} \label{backgr: refact-improv-design}

Our study included 32 undergraduate students currently pursuing their degrees in the Computer Science field. We considered our subjects as ``novices'' in Python because they reported having on average seven months of experience in Python, the language in which the programs were written. In general, the subjects had a minimum of six months and a maximum of 42 months of experience with programming languages, including Python, Java, JavaScript, C, and C\texttt{++}. They were recruited from three universities in Brazil, mainly through in-person invitations or text messages. All subjects were Brazilian Portuguese speakers enrolled in academic institutions.

Regarding the sample size, we performed a calculation considering the desired effect, significance level, and statistical power. The goal was to ensure a minimum power of 0.8, with a significance level of 0.05, using the t-test sample size calculation. Our analysis indicated that 26 subjects in two groups would be required to meet these criteria. Alternatively, given that we had 32 subjects instead of 26, our study can identify a moderate effect size of 0.5, maintaining a statistical power of 0.8 and a significance level of 0.05. A larger sample size provides sufficient sensitivity to detect a slightly smaller effect while maintaining statistical robustness.

We conducted the experiment at three locations to gather more subjects and to have a variety of subjects from different higher education institutions. However, different locations may influence subjects' visual attention. To mitigate this, we carefully organized the rooms to have similar conditions. For instance, the rooms were quiet, with minimal distractions, similar temperatures, and artificial light sources. We documented which subject performed the experiment at each location to account for potential differences.

\subsection{Design}\label{backgr: config-systems}

\begin{figure}[]%
    \centering
    \qquad
    {{\includegraphics[width=8.5cm]{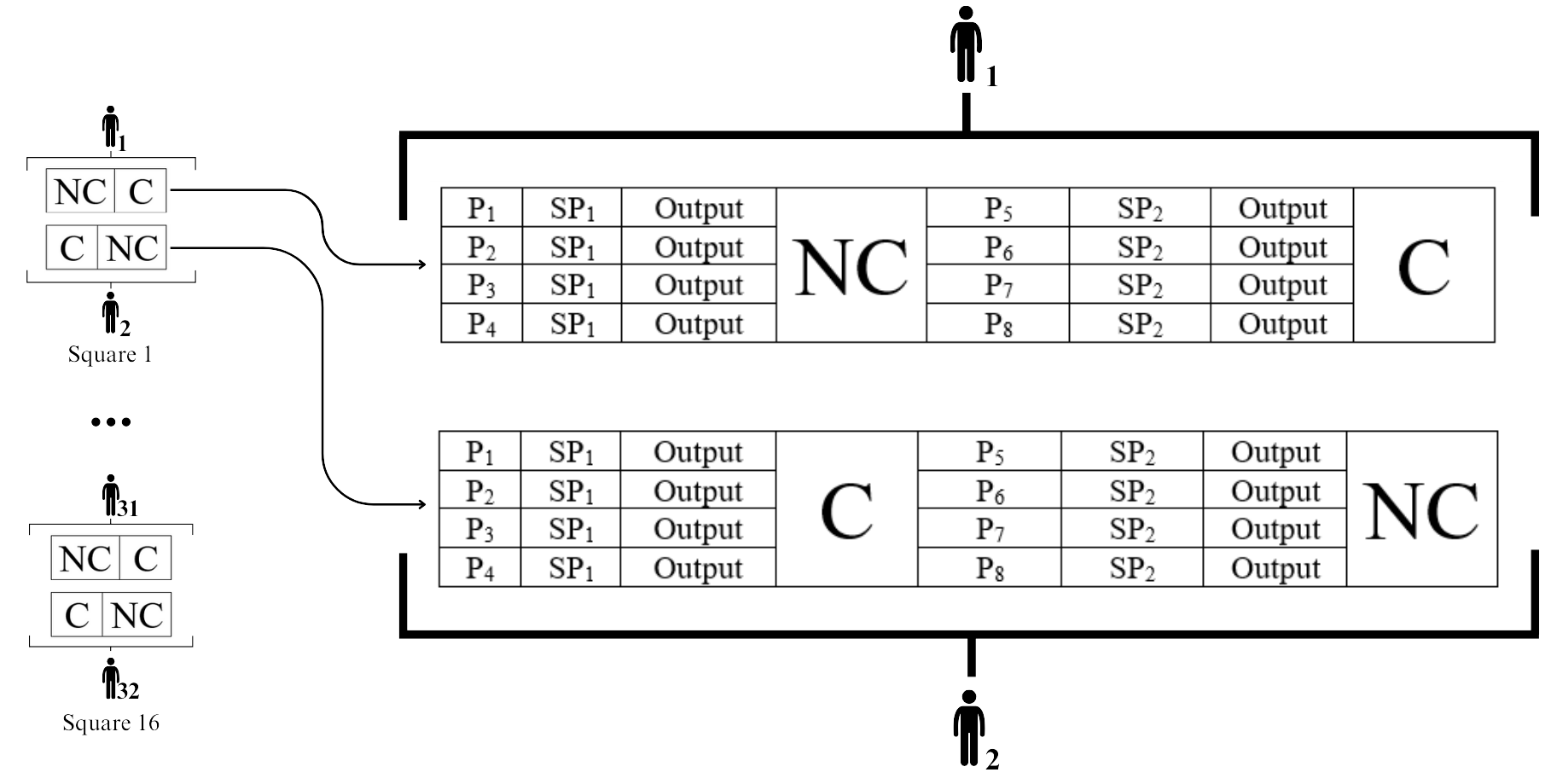} }}%
    \caption{Latin Square Structure. Each subject received four programs (P$_1$-P$_4$), which were \textit{PEP8 compliant} (C). These programs belonged to Program Set 1 (SP$_1$). Additionally, the subject received four programs (P$_5$-P$_8$) from Program Set 2 (SP$_2$), comprising \textit{PEP8 non-compliant} (NC).}%
    \label{fig: latin}%
\end{figure}

As illustrated in Figure~\ref{fig: latin}, each subject analyzed eight programs (P$_1$-P$_8$). To mitigate learning effects, we employed the Latin Square design~\cite{BO05ST}. Sixteen different programs were designed, divided into two sets of programs (SP$_1$ and SP$_2$). A subject analyzed four programs from set SP$_1$ and four programs from set SP$_2$. Another subject analyzed four programs from set SP$_1$ and four programs from set SP$_2$. Programs within the same set, although having different code programs, resulted in the same output. In all programs, subjects were required to specify the correct output, with no multiple-choice options. Given the program's input, subjects had to perform tasks such as calculating operations, and summing lists, among others.

\subsection{Evaluated PEP8 Style Guide Guidelines}\label{backgr: c-preproc}

We evaluated four guidelines from the PEP8 style guide (see  Table~\ref{tab:pep8-guidelines}): \textit{Whitespace}, \textit{Line Break Before Operator}, \textit{Multiple Statements on the Same Line}, and \textit{Comparison to True}. Following the guide's suggestions, we referred to the snippets as \textit{PEP8 compliant} and \textit{PEP8 non-compliant}.
\revision{
We selected PEP8 guidelines suitable for code snippets for novices in Python. 
}

\begin{table}[ht]
    \small
    \caption{PEP8 guidelines evaluated in this study.}
    \label{tab:pep8-guidelines}
    \begin{tabular}{p{1.3cm}p{2.8cm}p{3.0cm}}
        \toprule
        \textbf{Guideline} & \textbf{PEP8 compliant} &  \textbf{PEP8 non-compliant} \\
        \midrule
        White-\ space & \begin{tabular}[t]{@{}l@{}}\texttt{hypot2 = x*x + y*y}\\\texttt{c = (a+b) * (a-b)}\end{tabular} &  \begin{tabular}[t]{@{}l@{}}\texttt{hypot2 = x * x + y * y}\\\texttt{c = (a + b) * (a - b)}\end{tabular} \\
        \midrule
        Line Break Before Operator & \begin{tabular}[t]{@{}l@{}}\texttt{income = (gross\_wages}\\ \texttt{+ taxable}\\ \texttt{+ dividends}\\ \texttt{- ira\_deduction}\\ \texttt{- student)}\end{tabular} &  \begin{tabular}[t]{@{}l@{}}\texttt{income = (gross\_wages +}\\ \texttt{taxable +}\\ \texttt{dividends -}\\ \texttt{ira\_deduction -}\\ \texttt{student)}\end{tabular} \\
        \midrule
        Multiple Statements on the Same Line & \begin{tabular}[t]{@{}l@{}}\texttt{while t < 10:}\\\texttt{   t = delay()}\end{tabular} &  \begin{tabular}[t]{@{}l@{}}\texttt{while t < 10: t = delay()}\end{tabular} \\
        \midrule
        Comparison to True & \texttt{if greeting:} &  \texttt{if greeting == True:} \\
        \bottomrule
    \end{tabular}
\end{table}

The selected code snippets and guidelines for analysis were carefully chosen with a focus on the target demographic's background knowledge, the novice Python students. We have considered that Python served as the primary language covered in the Algorithms or Structured Programming courses at the universities where our experiment was conducted. Our objective was to explore and evaluate guidelines that align with the prevalent coding styles adopted by novice Python learners in these academic settings.

\subsection{Programs}
\label{backgr: undisc}

\begin{figure}[]%
    \centering
    \qquad
    {{\includegraphics[width=9cm]{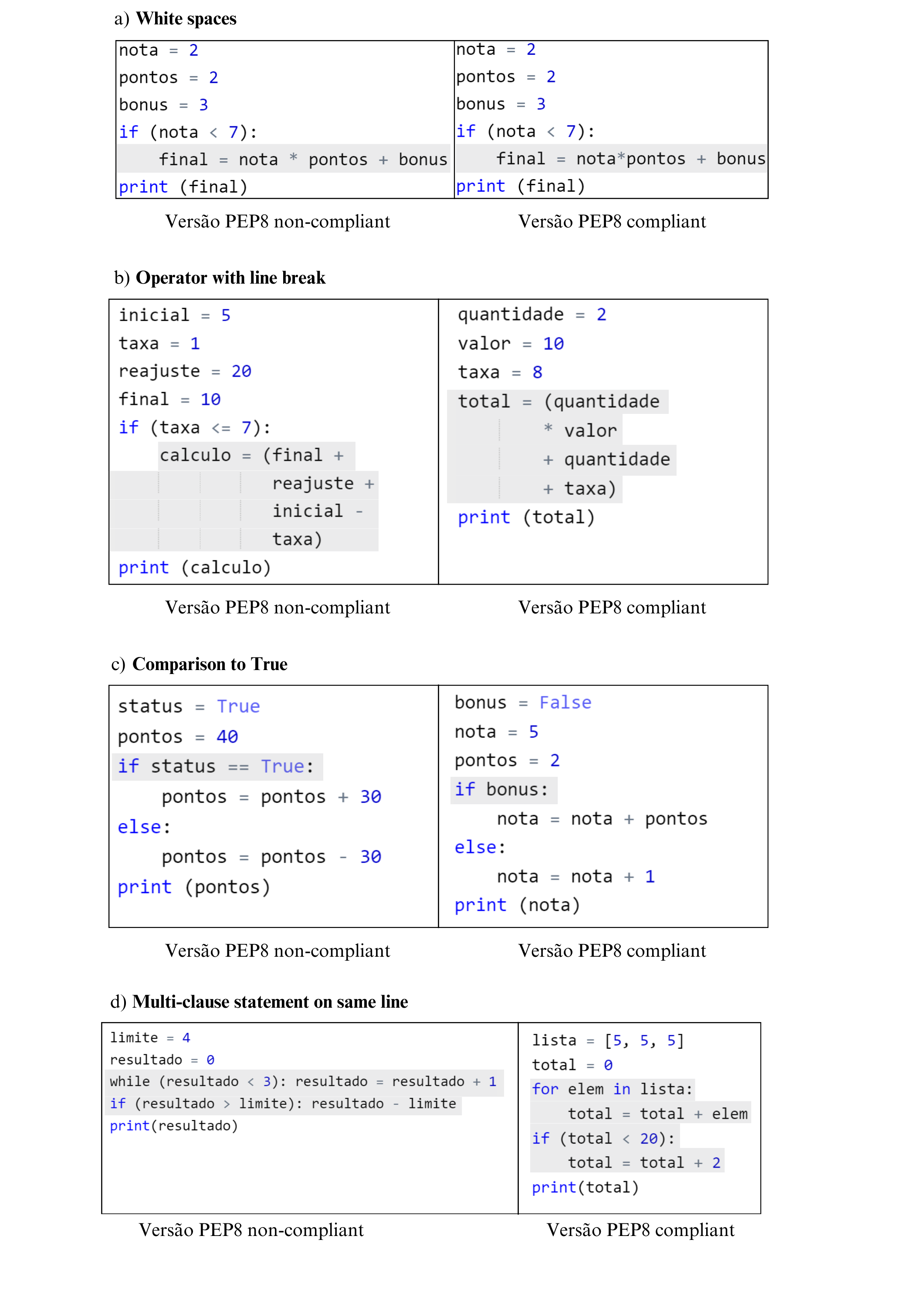} }}%
    \caption{Programs presented to the participants.}%
    \label{fig: programs}%
\end{figure}

We selected code snippets from repositories such as GeekForGeeks\footnote{\textcolor{blue}{https://www.geeksforgeeks.org/}} and Leetcode\footnote{\textcolor{blue}{https://leetcode.com/}}, for introductory programming activities. We prioritized problems with up to 11 lines of code, adapted for camera constraints, as illustrated in Figure~\ref{fig: programs}. Following a common methodology employed by code comprehension experiments~\cite{Oliveira2020evaluating}~\cite{Fakhoury2020}~\cite{Sharafi2021}, we asked subjects to predict the correct output of code snippets, without syntactic errors, addressing the four PEP8 guidelines evaluated in this study (Table ~\ref{tab:pep8-guidelines}).

The programs, displayed in Consolas 16, underwent a careful approach. Each program, whether the \textit{PEP8 compliant} or \textit{PEP8 non-compliant} version, had a single instance of one of the PEP8 style guide patterns. With a Latin Square design, no subject saw both the \textit{PEP8 compliant} or \textit{PEP8 non-compliant} versions of the same program.

In Figure~\ref{fig: programs}, we present examples of the programs used in the experiment, with one version containing the \textit{PEP8 non-compliant} pattern and another version containing the \textit{PEP8 compliant} pattern of PEP8 for the four styles evaluated. In Figure ~\ref{fig: programs}, we depict a set of programs with \textit{PEP8 non-compliant} versions (left-hand side) of the code, including guidelines such as Whitespace, Line Break Before Operator, Comparison to True, and Multiple Statements on the Same Line. The shaded areas indicate Areas of Interest (AOIs), corresponding to lines of code where the \textit{PEP8 non-compliant} and \textit{PEP8 compliant} versions differ.
We chose to use small-sized programs, with up to 11 lines of code, to fit the code on the screen. This choice may limit the applicability to more extensive programs. 


\subsection{Eye Tracking System}\label{backgr: refactoring}

Our research employed the Tobii Eye Tracker 4C with a sampling rate of 90 Hz. The eye tracker calibration followed the standard procedure, involving gazes at five calibration points, twice, and verification with eight points. The device was mounted at a distance of 50-60 cm from the subject on a laptop screen. Code tasks were displayed in full-screen mode, without the use of an Integrated Development Environment (IDE). We calculated a precision error of 0.7 degrees from this distance. For eye gaze analysis and metric collection, we developed a Python script.
\revision{
We used camera settings used in previous studies~\cite{da2023seeing,costa2021evaluating} on code comprehension using eye-tracking cameras.
We used a Dispersion-Based algorithm to classify the fixations. In particular, we used the Dispersion-Threshold Identification~\cite{salvucci2000identifying}. We also classified gaze samples as belonging to a fixation if the samples are located within a spatial region of approximately 0.5 degrees~\cite{nystrom2010adaptive}. 
We also implemented a simple Python script to create diagrams from data points using open source libraries to draw arrows and images, and create heatmaps. 
}

\subsection{Study Pilot} \label{backgr: refactoring-code}

We conducted pilot studies with four participants to refine materials and adjust the experiment's design, excluding these subjects from the final analysis. The process allowed us to simplify the programming tasks and focus exclusively on the impact of PEP8 guidelines on the codes, identifying and eliminating other variables that could influence the results. 

The study material included a collection of programs, a questionnaire for characterizing participants, and semi-structured interview questions. Program snippets were sourced from the PEP8 guide and introductory programming course datasets. Various aspects like code difficulty, font size, style, spacing, and indentation were evaluated. Tasks generally took under two minutes to complete, and questionnaire questions were refined. Identifiers were carefully chosen to convey information, such as using abbreviations like \texttt{elem} and specific terms like \texttt{bonus} for context clarity.

%% file: source/5-results.tex
\section{Results and Discussion}
\label{results-discussion}

In Table~\ref{resultstable}, we summarize the quantitative results of the collected metrics for each guideline with the statistical analysis. We present two perspectives of the metrics evaluated in this work, one examining only the AOI and the other examining the code as a whole. While time in the code, for example, consists of the time needed to examine and solve the task, regardless of the fixations made, time in the AOI consists of the time dedicated to examining only the region containing a style following or not following the PEP8 guidelines. 

Table~\ref{resultstable} also presents the data for the \textit{PEP8 non-compliant} version (column NC) and \textit{PEP8 compliant} version (column C) on the metrics highlighted in the second column. In the PD\% column, we present the percentage difference between the two versions concerning the particular metric. The percentage was calculated with respect to the NC version with an arrow that indicates how much the C version increased or decreased this percentage compared to the NC version. The NC and C columns are based on the median as a measure of central tendency, except for less sensitive attempts which are based on the mean. While time in the code represents the total effort to examine and solve the task, time in the AOI offers more specific insights, focusing exclusively on the region relevant to the PEP8 guideline. 

Concerning our RQ$_1$, the \textit{PEP8 compliant} version of the PEP8 guideline \textit{Line Break Before Operator} resulted in a reduction by 48.7\% in time in the AOI and 35.27\% in total time compared to the NC version. This suggests that following the PEP8 guideline can optimize the time spent on code analysis. It also highlights the importance of considering not only the total duration but also the efficiency in the code analysis. Notably, this distinction between time in the code and time in the AOI is vital when evaluating the NC and C versions, as it provides a more refined perspective on how following the PEP8 guidelines can influence not only the total time invested but also the efficiency and accuracy in the code analysis. This difference between the NC and C versions can be critical for understanding the overall impact of coding practices according to the established guidelines.

By following the \textit{Line Break Before Operator} guideline, Table~\ref{resultstable} highlights a decrease by 37.14\% in Fixation Duration, our RQ$_3$. This result points to a reduction in visual effort concentrated in the area associated with the evaluated pattern when PEP8 guidelines are followed compared to the NC version. Additionally, when analyzing the metrics of Horizontal Regressions and Vertical Regressions for the \textit{White Space} guideline in our \revision{RQ$_4$}, we observe that following the PEP8 guidelines is associated with a 25\% reduction in Horizontal Regressions, indicating a smoother and continuous reading.

\setlength{\tabcolsep}{2pt} 
\renewcommand{\arraystretch}{0.8} 
\begin{table*}[htbp]
\centering
\caption{Summary of metrics: time, submissions, fixation duration, fixation count, horizontal regressions, and vertical regressions for each PEP8 guideline. NC represents the \textit{PEP8 non-compliant} version and C the \textit{PEP8 compliant} version; PD represents the percentage difference; n/a represents unassigned value. Bold font represents statistically significant differences. \label{resultstable}}
\begin{tabular}{|c|l|r|r|r|r|r|r|r|r|r|r|}
\hline
\rowcolor{gray!25}
\multicolumn{1}{|c|}{\multirow{1}{*}{}} & \multicolumn{1}{|c|}{\multirow{1}{*}{}}& \multicolumn{5}{c|}{\textbf{In the AOI}} & \multicolumn{5}{c|}{\textbf{In the Code}} \\ 

\cline{3-12} 
\rowcolor{gray!25}
\multicolumn{1}{|c|}{\shortstack{\textbf{PEP8 Guidel.}}} & \multicolumn{1}{|c|}{\shortstack{\textbf{Metrics}}} & \multicolumn{1}{c|}{NC} & \multicolumn{1}{c|}{C} & \multicolumn{2}{c|}{PD\%} & \multicolumn{1}{c|}{\textit{p}-val.} & \multicolumn{1}{c|}{NC} & \multicolumn{1}{c|}{C} & \multicolumn{2}{c|}{PD\%} & \multicolumn{1}{c|}{\textit{p}-val.}\\ 

\hline

\multirow{8}{*}{\begin{tabular}[c]{@{}c@{}}\textbf{White} \\ \textbf{Space}\end{tabular}} & Time & 6 & 5.69 & \multicolumn{2}{c|}{↓5.25} & 0.82 & 20 & 18.91 & \multicolumn{2}{c|}{↓5.45} & 0.63\\ \cline{2-12}
 & Submissions & n/a & n/a  & \multicolumn{2}{c|}{n/a}  & n/a & 1 & 1 & \multicolumn{2}{c|} 0 & 0.36 \\ \cline{2-12}
 & Fixations Duration & 10 & 9.5 & \multicolumn{2}{c|}{↓5} & 0.86 & 23.5 & 27 & \multicolumn{2}{c|}{↑12.7} & 0.97 \\ \cline{2-12}
 & Fixations Count & 2.92 & 2.96 & \multicolumn{2}{c|}{↑1.54} & 0.7 & 7.17 & 8.59 & \multicolumn{2}{c|}{↑19.73} & 0.91 \\ \cline{2-12}
 & \begin{tabular}[c]{@{}l@{}}Horizontal \\ Regressions\end{tabular} & 2 & 1.5 & \multicolumn{2}{c|}{↓25} & 0.46 & 4 & 4 & \multicolumn{2}{c|}{0.86} & 0.86 \\ \cline{2-12}
 & \begin{tabular}[c]{@{}l@{}}Vertical \\ Regressions\end{tabular} & 0 & 0 & \multicolumn{2}{c|}{n/a} & n/a & 6 & 7 & \multicolumn{2}{c|}{↑8.33} & 0.96\\ \hline

\multirow{8}{*}{\begin{tabular}[c]{@{}c@{}}\textbf{Line}\\ \textbf{Break} \\ \textbf{Before} \\ \textbf{Operator}\end{tabular}} & Time & 14.27 & 7.32 & \multicolumn{2}{c|}{↓48.7} & \textbf{0.002} & 30.7 & 19.8 & \multicolumn{2}{c|}{↓35.27} & \textbf{0.005}\\ \cline{2-12}
 & Submissions & n/a & n/a & \multicolumn{2}{c|}{n/a} & 0.0 & 1.13 & 1 & \multicolumn{2}{c|}{↓11.11} & \textbf{0.04} \\ \cline{2-12}
 & Fixations Duration & 17.5 & 11 & \multicolumn{2}{c|}{↓37.14} & \textbf{0.02} & 36 & 31.5 & \multicolumn{2}{c|}{↓12.5} & 0.11 \\ \cline{2-12}
 & Fixations Count & 5.46 & 4.21 & \multicolumn{2}{c|}{↓22.8} & \textbf{0.01} & 11.06 & 10.11 & \multicolumn{2}{c|}{↓8.63} & \textbf{0.04} \\ \cline{2-12}
 & \begin{tabular}[c]{@{}l@{}}Horizontal \\ Regressions\end{tabular} & 6 & 2 & \multicolumn{2}{c|}{↓66.66} & \textbf{0.03} & 6 & 4 & \multicolumn{2}{c|}{↓27.27} & \textbf{0.02} \\ \cline{2-12}
 & \begin{tabular}[c]{@{}l@{}}Vertical \\ Regressions\end{tabular} & 2.5 & 1 & \multicolumn{2}{c|}{↓60} & \textbf{0.03} & 12 & 85 & \multicolumn{2}{c|}{↓29.16} & \textbf{0.007} \\ \hline

\multirow{8}{*}{\begin{tabular}[c]{@{}c@{}}\textbf{Multiple}\\ \textbf{Statements} \\ \textbf{on the same} \\ \textbf{Line}\end{tabular}} & Time & 28.95 & 22.89 & \multicolumn{2}{c|}{↓20.92} & 0.2 & 42.9 & 32.5 & \multicolumn{2}{c|}{↓24.16} & 0.06 \\ \cline{2-12}
 & Submissions & n/a & n/a & \multicolumn{2}{c|}{n/a} & {n/a} & 1.5 & 1.13 & \multicolumn{2}{c|}{↓25} & 0.1 \\ \cline{2-12}
 & Fixations Duration & 46.5 & 32.5 & \multicolumn{2}{c|}{↓30.1} & 0.36 & 62 & 45 & \multicolumn{2}{c|}{↓27.41} & 0.21 \\ \cline{2-12}
 & Fixations Count & 14.99 & 11.58 & \multicolumn{2}{c|}{↓27.74} & 0.37 & 20.04 & 15.58 & \multicolumn{2}{c|}{↓22.24} & 0.15 \\ \cline{2-12}
 & \begin{tabular}[c]{@{}l@{}}Horizontal \\ Regressions\end{tabular} & 13 & 14 & \multicolumn{2}{c|}{↑7.69} & 0.76 & 13 & 10 & \multicolumn{2}{c|}{↓24} & 0.33 \\ \cline{2-12}
 & \begin{tabular}[c]{@{}l@{}}Vertical \\ Regressions\end{tabular} & 2.5 & 4 & \multicolumn{2}{c|}{↑60} & 0.63 & 11 & 9.5 & \multicolumn{2}{c|}{↓13.63} & 0.8 \\ \hline

\multirow{8}{*}{\begin{tabular}[c]{@{}c@{}}\textbf{Comparison}\\ \textbf{to True}\end{tabular}} & Time & 2.46 & 2.24 & \multicolumn{2}{c|}{↓8.94} & 0.87 & 13.6 & 15.9 & \multicolumn{2}{c|}{↑16.84} & 0.05 \\ \cline{2-12}
 & Submissions & n/a & n/a & \multicolumn{2}{c|}{n/a} & {n/a} & 1 & 1 & \multicolumn{2}{c|}{0} & {n/a} \\ \cline{2-12}
 & Fixations Duration & 3 & 3.5 & \multicolumn{2}{c|}{↑16.66} & 0.56 & 18 & 22 & \multicolumn{2}{c|}{↑22.22} & 0.28 \\ \cline{2-12}
 & Fixations Count & 1.11 & 1.03 & \multicolumn{2}{c|}{↑8.25} & 0.31 & 7.02 & 6.21 & \multicolumn{2}{c|}{↑13.12} & 0.16 \\ \cline{2-12}
 & \begin{tabular}[c]{@{}l@{}}Horizontal \\ Regressions\end{tabular} & 0.5 & 0 & \multicolumn{2}{c|}{0} & 0.05 & 2 & 3 & \multicolumn{2}{c|}{↑25} & 0.16 \\ \cline{2-12}
 & \begin{tabular}[c]{@{}l@{}}Vertical \\ Regressions\end{tabular} & 0 & 0 & \multicolumn{2}{c|}{n/a} & {n/a} & 5 & 6 & \multicolumn{2}{c|}{↑20} & 0.15 \\ \hline
\end{tabular}
\end{table*}

Concerning the \textit{Comparison to True} guideline, Table~\ref{resultstable} presents important nuances regarding the influence of PEP8 guidelines on various metrics. Although following this guideline resulted in an 8.94\% reduction in time in the AOI, it showed an increase of 16.84\% in time in the code (RQ$_1$). There is also a simultaneous increase of 16.66\% in Fixation Duration and 8.25\% in Fixation Count (RQ$_3$). These results suggest a potential trade-off between overall efficiency and detail in the code analysis, indicating that, while following the PEP8 guideline may speed up the analysis, there may be a cost associated with the frequency and duration of revisions.

Additionally, we observed that in general, there were no significant differences in time spent and the number of answer attempts between the different PEP8 patterns for the \textit{White Space} \textit{Multiple Statements on the same Line} and \textit{Comparison to True} criteria. However, there was a noteworthy decrease in fixation duration and count for the \textit{Line Break Before Operator} and \textit{Multiple Statements on the same Line} patterns, indicating enhanced efficiency in understanding and processing these patterns. Moreover, a decrease in the number of horizontal and vertical regressions was observed for these same patterns, suggesting a more linear and organized reading of the code. Interestingly, certain metrics, such as time and fixation count, exhibited performance differences between the PEP8 patterns compared to the control group \textit{Comparison to True}, suggesting variations in the ease of understanding and processing the different patterns. 
Next we discuss our results in more details.

\subsection{Preferences of the Subjects }
\label{results:preferences}

\begin{figure}[]%
    \centering
    \qquad
    {{\includegraphics[width=\linewidth]{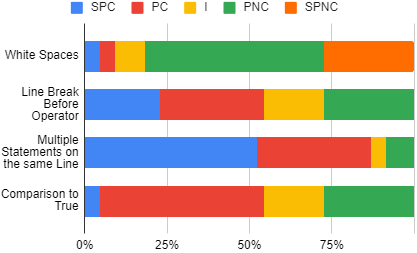} }}%
    \caption{Subject's preferences for the \textit{PEP8 compliant} and \textit{PEP8 non-compliant} versions of the PEP8 guidelines. We used the following acronyms: Strongly Prefers \textit{PEP8 compliant} (SPC); Prefers \textit{PEP8 compliant} (PC); Indifferent (I); Prefers \textit{PEP8 non-compliant} (PNC); Strongly Prefers \textit{PEP8 non-compliant} (SPNC).
    }%
    \label{fig:ent}%
\end{figure}

After solving all programming tasks, we showed the \textit{PEP8 compliant} and \textit{PEP8 non-compliant} versions, side by side, for each PEP8 guideline, to the subjects. The differences in the code snippets between the \textit{PEP8 compliant} and \textit{PEP8 non-compliant} versions were highlighted. Figure~\ref{fig:ent} depicts the subjects' overall responses, indicating their preferences among the presented versions, expressing whether they had a Strong Preference, Preference, or were Indifferent to the versions presented: Strongly Prefers (SP) \textit{PEP8 non-compliant} (SPNC) or \textit{PEP8 compliant} (SPC), Prefers (P) \textit{PEP8 non-compliant} (PNC) or \textit{PEP8 compliant} (PC), and Indifferent (I).

Subjects predominantly preferred or strongly preferred the \textit{PEP8 non-compliant} version for the \textit{White Spaces} guideline. However, concerning the guideline for \textit{Multiple Statements on the Same Line}, the majority preferred or strongly preferred the \textit{PEP8 compliant} version. For \textit{Line Break Before Operator} and \textit{Comparison to True}, the majority preferred the \textit{PEP8 compliant} version with PEP8 guidelines.
This response pattern may suggest that while some PEP8 guidelines are widely accepted, others are considered less critical or subject to personal interpretation.


In addition to expressing preferences among the presented versions, the subjects were asked about the reasons behind their choices. The more in-depth analysis of these interview data is enriched through triangulation with eye tracking metrics. The discussion of this triangulation is presented in the subsequent sections as we investigate each pattern of the PEP8 guidelines.

\subsection{PEP8 Guidelines} \label{results:guidelines}

In this section, we triangulate and discuss each guideline evaluated, starting with \textit{White Space} (see Section~\ref{results:whitespace}), followed by \textit{Line Break Before Operator} (see Section~\ref{results:linebreak}), \textit{Multiple Clauses} (see Section~\ref{results:mclauses}), and finally, \textit{Comparison to True} (see Section~\ref{results:comparisonTrue}).

\subsubsection{White Space} \label{results:whitespace}

Regarding our RQ$_1$, it was observed that subjects spent more time in the AOI in the \textit{PEP8 non-compliant}. Apparently, adding space between the multiplication operator also caused subjects to regress more in the \textit{PEP8 non-compliant} version. Not following the PEP8 guideline impacted the number of horizontal eye regressions for the novices, reducing it by approximately 25\%. To clarify this aspect further, we will discuss how eye regression can indicate issues in code readability with the \textit{PEP8 non-compliant} version of the PEP8 recommendation for \textit{White Space}.

\begin{figure}[]%
    \centering
    \qquad
    {{\includegraphics[width=\linewidth]{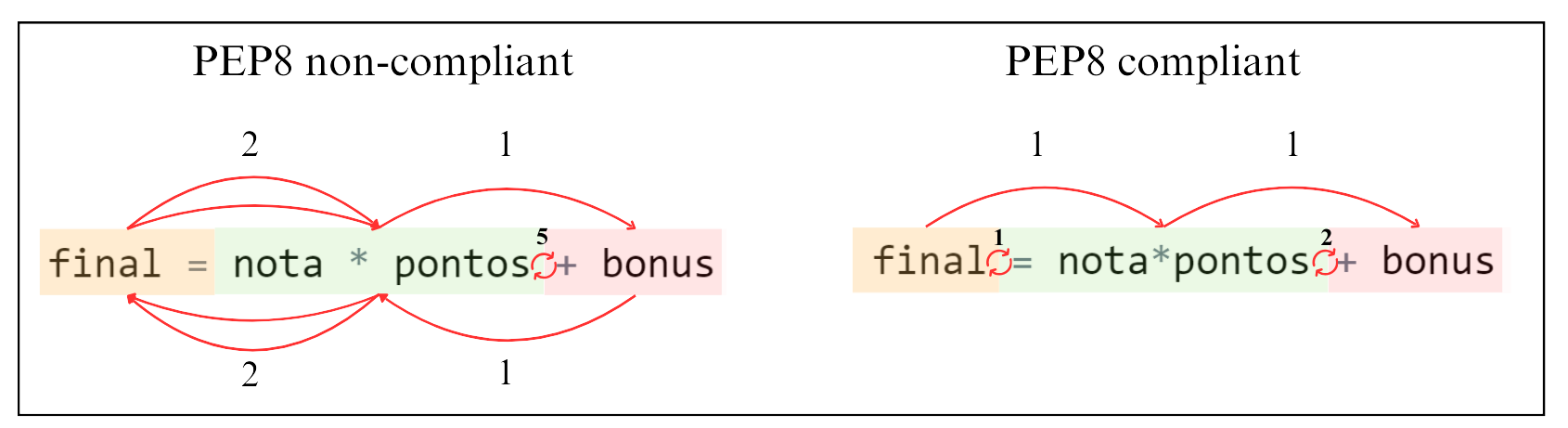} }}%
    \caption{Eye \textit{regression} and \textit{progression} of reading in the AOI of the \textit{White Space} pattern for the \textit{PEP8 non-compliant} version (left-hand side) and \textit{PEP8 compliant} version (right-hand side) of a subject. The green region contains the portion of the code where the PEP8 pattern has been applied or not. Arrows pointing from right to left represent reading progression, and from left to right, eye returning to the previous region of the code. The numbers on the arrows represent the number of times progression or regression occurred.
    }%
    \label{fig:transicaows}%
\end{figure}

Still concerning our \revision{RQ$_4$}, it was noted that the \textit{PEP8 non-compliant} version exhibited a slightly higher number of horizontal regressions in the AOI. From data extracted from one of the subjects in the experiment, Figure~\ref{fig:transicaows} presents the horizontal regressions captured by this subject's eye tracking camera. On the left-hand side of Figure~\ref{fig:transicaows}, we have the code snippet containing the \textit{PEP8 non-compliant} version of PEP8 guideline \textit{White Space} (Table~\ref{resultstable}), and on the right-hand side, the version with the \textit{PEP8 compliant} guideline. Each line between code blocks indicates either a code \textit{regression}, where the gaze returns to a previous section of code, or \textit{progression}, where the gaze moves on to a further section. The spacing around the multiplication operator (seen on the left-hand side of Figure ~\ref{fig:transicaows}) may account for an increase in horizontal regressions, potentially indicating a greater visual effort and reading repetition. 

\begin{figure}[]%
    \centering
    \qquad
    {{\includegraphics[width=\linewidth]{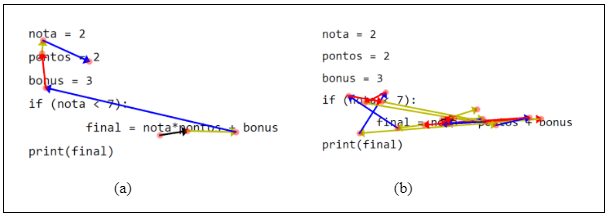} }}%
    \caption{Sequential gaze transitions when reading a program with the \textit{White Space} pattern for the \textit{PEP8 non-compliant} (a) and \textit{PEP8 compliant} (b) versions of different subjects.
    }%
    \label{fig:transicaows2}%
\end{figure}

Figure~\ref{fig:transicaows2} showcases two code excerpts marked with gaze transitions from different subjects, one with the \textit{PEP8 non-compliant} version (a) and the other with the \textit{PEP8 compliant} version (b), according to PEP8's \textit{White Space} standard. The color-coded lines track the sequence of eye movements across the code versions. It is observed that the \textit{PEP8 compliant} version (b) had fewer regressions, suggesting that code adhering to the PEP8 \textit{White Space} guideline is easier to follow and understand, as opposed to the \textit{PEP8 non-compliant} version (a), where more frequent regressions occurred, especially in the section not following the PEP8 \textit{White Space} guideline, a pattern repeated across subjects.

Regarding the subjects' preferences, approximately 81\% of the subjects preferred the \textit{PEP8 non-compliant} version of PEP8 guide for \textit{White Space}. When we asked why they preferred one style over the other, one of the subjects responded as follows: ``...I don't like the idea of [the operator followed by the variable without using space], I think the lack of spacing is confusing...''. The absence of space between the multiplication operator displeases some subjects, which may justify the preference for the \textit{PEP8 non-compliant} pattern. However, we can conclude that although the majority of subjects preferred the \textit{PEP8 non-compliant} pattern, it negatively impacted code readability when considering overall time and eye regression data. Interestingly, participants' comments on specific tasks indicate that, despite a preference for more spaces, understanding of the code was not necessarily improved by this practice. 

\subsubsection{Line Break Before Operator}
\label{results:linebreak}

For the \textit{Line Break Before Operator} guideline, the PEP8 guide provides the following justification: ``the eye has to do extra work to figure out which items are added and which are subtracted''~\cite{van2001pep}. In our study, we found that the \textit{PEP8 non-compliant} version  reduced the performance of the subjects in nearly all evaluated metrics, as shown in Table~\ref{resultstable}, corroborating the findings of Rossum et al.~\cite{van2001pep}. Statistically, the version recommended by PEP8 for \textit{Line Broke Before Operator} demonstrated better performance. Regarding our \revision{RQ$_4$}, there is indeed an indication of more vertical and horizontal regressions for the \textit{PEP8 non-compliant} guideline, supporting PEP8's assertion.

\begin{figure}[]%
    \centering
    \qquad
    {{\includegraphics[width=\linewidth]{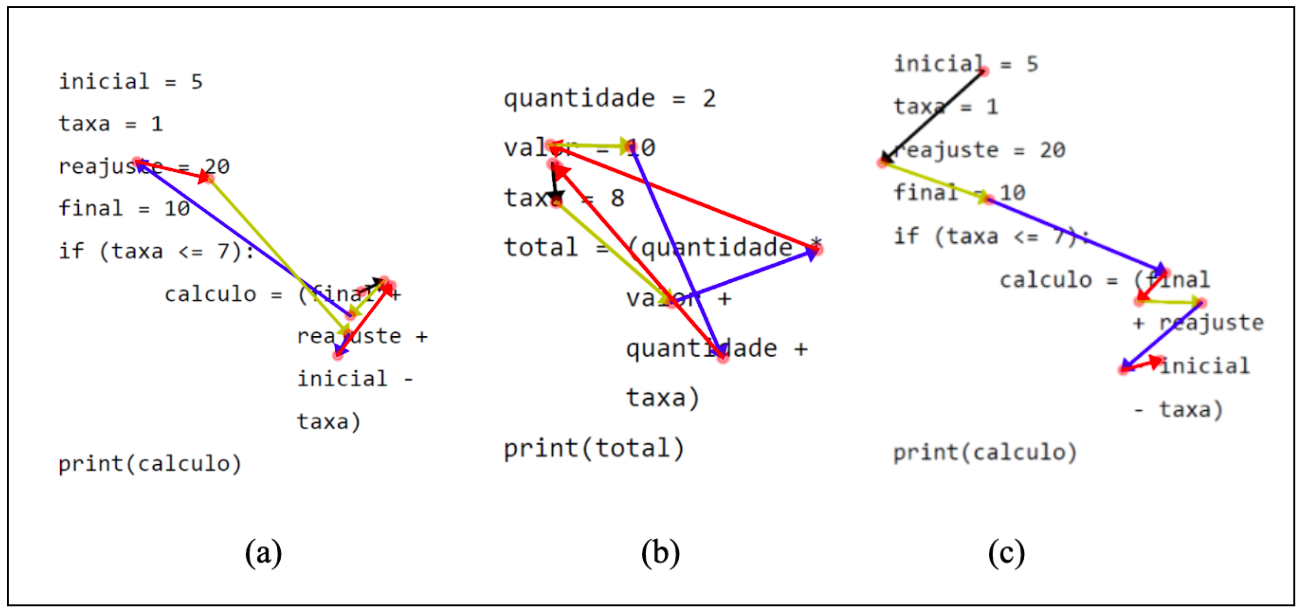} }}%
    \caption{Eye transition for \textit{Line Break Before Operator} guideline from three subjects, with code snippets (a) and (b) representing the \textit{PEP8 non-compliant} version of the guideline and code (c) representing the \textit{PEP8 compliant} version.
    }%
    \label{fig:transicaoLineBreak}%
\end{figure}

In Figure~\ref{fig:transicaoLineBreak}(a) and (b), we illustrate the eye transition where two subjects regress their gaze to the operator at a certain point in the program reading, suggesting an extra visual effort to return to the operator (PEP8 non-compliant version). This behavior was also observed in the data from other subjects. However, in the snippet from Figure~\ref{fig:transicaoLineBreak}(c) with the \textit{PEP8 compliant} version, one subject demonstrates a sequential reading flow without regressing, meaning they do not go back to the operator. This observation was noted in the data from other subjects for the \textit{PEP8 compliant} version of the pattern discussed in this subsection.

For the \textit{Line Break Before Operator} guideline, the majority prefers the \textit{PEP8 compliant} version. The main reasons were ``left alignment'' and ``ease of reading'', stating that they are more accustomed to a left-aligned format for mathematical operations. This observation complements the discussion on the preference for line breaks before or after a binary operator, which revealed divided opinions among participants. Some found that breaking lines facilitated understanding by visually separating the components of complex calculations, despite not being accustomed to this formatting. Others, however, expressed that this approach caused estrangement and preferred the presentation of calculations in a single line, arguing that this made the sequence of operations more straightforward and easier to follow. This divergence of opinions highlights how personal familiarity significantly influences the perception of code readability, with some valuing the clarity provided by line breaks in extensive operations, while others see it as a barrier to the immediate understanding of mathematical operations.

\subsubsection{Multiple Clauses on the Same Line} \label{results:mclauses}


Answering our RQ$_2$, the number of submissions for the \textit{PEP8 non-compliant} version was slightly higher than for the \textit{PEP8 compliant} one regarding the guideline \textit{Multiple Statements on the Same Line}. To better understand it, we analyze one of the subjects' comments: ``...separating, we understand better what the if does because we can get confused if the \texttt{if} is inside the \texttt{while}...''. In this comment, it is possible to identify that the lack of indentation can cause confusion. According to this subject, there is confusion when one clause is followed by another on the same line.

\begin{figure*}[]%
    \centering
    \qquad
    {{\includegraphics[width=0.8\linewidth]{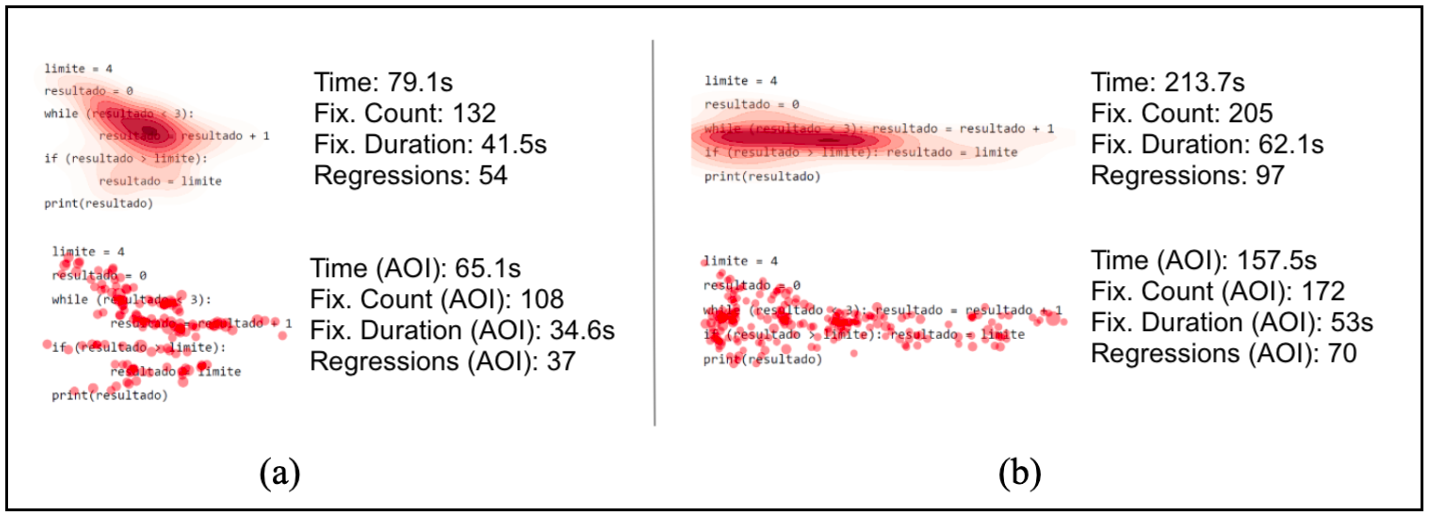} }}%
    \caption{Heatmaps and Fixation for the \textit{Multiple Clauses on the Same Line} pattern.
    }%
    \label{fig:heatmapMultiClauses}%
\end{figure*}

Concerning the fixations count, our \revision{RQ$_3$}, we observed that the \textit{PEP8 compliant} version reduced the number of fixations by up to 30\% and the duration of these fixations by up to 27\% in the AOI for the guideline discussed in this subsection, compared to the \textit{PEP8 non-compliant} pattern. From the heatmap and fixation count, it was possible to notice that, for two subjects, we were able to identify some nuances that may justify these data (Figure~\ref{fig:heatmapMultiClauses}).

The right-hand side (a) in Figure~\ref{fig:heatmapMultiClauses} demonstrates the code with the \textit{PEP8 compliant} version. Both code snippets on side (a) and on side (b) have the same algorithmic complexity. However, we noticed that the performance of the subject who solved the \textit{PEP8 non-compliant} version was worse in terms of time, our RQ$_1$, number of fixations and fixation duration, our RQ$_3$, in the AOI. The overall data presented in Table~\ref{resultstable} also reflects this difference.

Regarding preference, Figure~\ref{fig:ent} shows that almost all subjects prefer the \textit{PEP8 compliant} version of PEP8 for the guideline \textit{Multiple Clauses in the Same Line}. The reason reported by some Python novices was that in the \textit{PEP8 compliant} version, there is more clarity about whether the \texttt{if} statement is inside the \texttt{while} loop or not.

The strong preference for compliance with PEP8 underscores the importance of readability and clarity in code, reflecting familiarity with coding practices that prioritize these aspects. Participants found tasks easier due to the clear structure of the code, as recommended by PEP8, which facilitates comprehension even with multiple variables and operations. These well-defined and transparent standards improve code readability and assist beginners in learning programming.

\subsubsection{Comparison to True}
\label{results:comparisonTrue}

\begin{figure}[]%
    \centering
    \qquad
    {{\includegraphics[width=\linewidth]{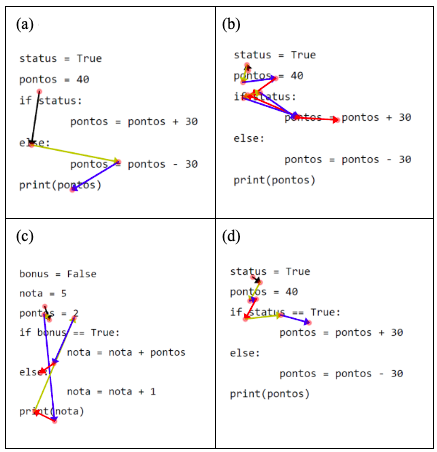} }}%
    \caption{Eye tracking transition for the \textit{Comparison to True} guideline. Versions (a) and (b) are \textit{PEP8 compliant}, while versions (c) and (d) are \textit{PEP8 non-compliant}, according to PEP8, representing four distinct subjects.
    }%
    \label{fig:transitionsComparisonTrue}%
\end{figure}

In Figure~\ref{fig:transitionsComparisonTrue}(a), we observed that a subject analyzed the \texttt{else} block in the code, which would not be an expected behavior, given that the value of \texttt{status} is \texttt{True}. In Figure~\ref{fig:transitionsComparisonTrue}(b), it can be noted that the subject returned several times to the \texttt{if} block, probably to check the value of \texttt{status}. One of the subjects even mentioned that the value of \texttt{status} without the comparison to \texttt{True} is not clear. This comment justifies the gaze behavior in the section related to \textit{status}.

\begin{figure*}[]%
    \centering
    \qquad
    {{\includegraphics[width=0.7\linewidth]{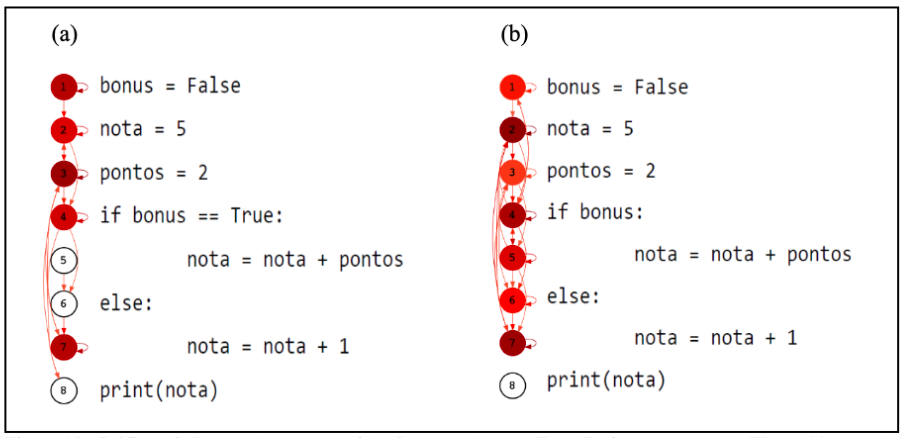} }}%
    \caption{Regression Graphs for the \textit{Comparison to True} guideline. Data for the \textit{PEP8 non-compliant} version (a) and \textit{PEP8 compliant} version (b) from two distinct subjects.
    }%
    \label{fig:transitionsGraph}%
\end{figure*}

Regarding the regression analyses \revision{(RQ$_4$)}, Figure~\ref{fig:transitionsGraph} depicts two graphs where the node represents a line of code (left and right-hand sides), and the edges indicate regressions, progressions, or returns to the same line of code. The \textit{PEP8 non-compliant} pattern shows a lower number of gaze returns compared to the \textit{PEP8 compliant} pattern. In fact, after checking the value of \texttt{bonus} in the \texttt{if} block (which is \texttt{False}), both in the \textit{PEP8 non-compliant} pattern, as represented in Figure~\ref{fig:transitionsGraph}(a), and in Figure~\ref{fig:transitionsComparisonTrue}(c), subjects direct their gaze to the \texttt{else} block in the code. This behavior is expected considering the value of \texttt{bonus}. However, this observation does not apply to the \textit{PEP8 compliant} pattern in Figure~\ref{fig:transitionsGraph}(b), as subjects explore code snippets that do not require verification. Furthermore, when comparing eye tracking data for the code snippets in Figure~\ref{fig:transitionsComparisonTrue}(b) and (d), the \textit{PEP8 non-compliant} version reveals that the subject's reading was more sequential and did not involve returns to check the value of \texttt{status}, as observed in the \textit{PEP8 compliant} pattern.

The \textit{PEP8 non-compliant} showed better performance, contradicting PEP8 guidelines. Participant feedback suggests that direct comparison of boolean values with \texttt{True} would be more appropriate, indicating performance improvement. This method simplifies boolean conditions, promoting readability and quick decision-making, especially useful for novices. PEP8 could revisit coding style guidelines.

%% file: source/6-threats.tex
\section{Threats to Validity}
\label{section:threats}

We conducted the experiment at three locations to gather more subjects and to have a variety of subjects from different higher education institutions. However, different locations may influence subjects' visual attention. To mitigate this, we carefully organized the rooms to have similar conditions. For instance, the rooms were quiet, with minimal distractions, similar temperatures, and artificial light sources. We documented which subject performed the experiment at each location to account for potential differences.

We allocated a total time of 40 minutes for each subject and assigned them eight programs, which could have influenced visual effort. To minimize this threat, we designed simple and short programs, each with only one instantiated pattern. Given the simplicity of the programs, most subjects solved them before the time limit. 

We chose to use small-sized programs, with up to 11 lines of code, to fit the code on the screen. This choice may limit the applicability to more extensive programs. However, previous works have employed code snippets with a similar number of lines, such as in Costa et al.~\cite{da2023seeing}. If we identify disparities in short snippets, we expect that longer segments may reveal more pronounced differences. Nevertheless, to support such expectations, it is imperative to conduct further studies with more extensive code snippets.

Within the scope of our study, we directed our attention to Python novices, limiting the generalization to more experienced developers in the language. 
\revision{However, using students as participants remains a valid simplification of reality needed in laboratory contexts~\cite{DBLP:journals/ese/FalessiJWTMJO18,DBLP:conf/icse/SalmanMJ15}.}
We plan to explore the same topic of this study with more experienced developers in the future.

%% file: source/7-related-work.tex
\section{Related Work}
\label{sec:related}
Previous approaches~\cite{bauer2019indentation,dos2018impacts} identified that certain coding styles negatively affect Java code readability, with whitespace being a particular concern. 
Our eye-tracking study revealed that in Python, following PEP8's \textit{White Space} guidelines may not align with better readability, as less spacing could potentially reduce visual regression and improve comprehension. 

Sharafi et al.~\cite{sharif2010aneye} examined the influence of Camel Case and underscore coding styles on code comprehension, measuring time, response correctness, and visual effort through eye tracking. They found a significant improvement in the time and visual effort with the underscore style. In a subsequent investigation~\cite{sharafi2012women} on the same styles, considering the gender of the subjects, no significant differences were identified in metrics such as time, accuracy, and visual effort. In our study, we sought similar metrics, namely time, number of attempts, and visual effort, however, in a different context.

Stefik and Siebert~\cite{stefik2013empirical} investigated how programming language syntax affects the comprehension of novice programmers. They asked novices to assess the intuitiveness of different programming language constructs and found that syntactic choices in commercial programming languages tend to be more intuitive for novices, influencing their initial programming accuracy rates. In contrast, our approach also considered the context of novices but specifically focused on a programming language. Additionally, we employed objective metrics to evaluate the performance of novices during code-solving tasks, providing a structured and quantitative analysis of code readability and how it could impact the accuracy of responses to proposed tasks containing the patterns.

Costa et al.~\cite{da2023seeing,CostaGheyi2023} used eye tracking to study how code patterns such as \textit{atoms of confusion} affect the understanding of Python code among novices. Their experiment with 32 participants showed that these atoms can complicate code comprehension, leading to increased effort and more attempts to solve coding problems. This research underscores the use of eye tracking in identifying which code patterns most hinder the novices' code comprehension.

\revision{Sharif et al.~\cite{sharafi2020practical} present a guideline to conduct eye tracking studies.} We followed rigorous guidelines for eye-tracking studies to examine how minor coding styles, as recommended by PEP8, affect Python code readability. Our findings reveal differences in readability and stylistic preferences, showcasing eye tracking's capability to measure the influence of coding patterns on both the performance and perception of programmers, thus highlighting its critical role in enhancing the understanding of coding practices and guiding future research.

%% file: source/8-conclusions.tex
\section{Conclusions}
\label{sec:conclusions}
In this work, we explore eye tracking as a method to gain insights into the guidelines of a code style guide. We conducted a controlled experiment with an eye tracker to assess the impact of four PEP8 guidelines on code readability, analyzing how the \textit{PEP8 compliant} and \textit{PEP8 non-compliant} versions, as defined by PEP8, affected the time, number of attempts, and visual effort of 32 Python novices.

We observed that non-adherence to PEP8's proper spacing increases time, number of fixations and horizontal regressions. Moreover, omitting line breaks before operators, against PEP8 recommendations, increased regressions, validating the importance of these practices for code readability.
The analysis of the \textit{Multiple Clauses on the Same Line} pattern showed that following the PEP8 guideline to separate clauses onto different lines reduced both the number and duration of fixations, enhancing code comprehension due to the clarity provided by this separation. Surprisingly, for the \textit{Comparison to True} pattern, results suggested that direct comparisons with Boolean values (True/False) were more effective, indicating that, in certain cases, deviating from PEP8 recommendations might actually aid in novice programmers' understanding of the code.


\revision{
For educators, we recommend paying close attention to the guidelines used in classes that impact undergraduate students' program comprehension. Based on our findings, explicitly using \texttt{== True} in conditions may help novices better understand the code. 
For practitioners, it is crucial to thoroughly understand coding style guidelines before adopting them. We recommend selecting guidelines supported by experimental evaluations rather than persuasive statements.
For researchers, we recommend evaluating more coding styles using robust methodologies to understand the impact of each style on code comprehension. It is also important to propose coding styles like PEP8 using proper methodologies.
The use of eye-tracking cameras can help researchers identify gaze transition patterns that could be integrated into more advanced IDEs. Advanced IDEs could use eye-tracking cameras to monitor developers' gaze transitions. These cameras, equipped with machine learning models and patterns of gaze transitions, could help automatically adjust code based on developers' preferences. 
}


For future research, we will explore the effects of lesser-studied PEP8 guidelines and their differing impacts on novice versus experienced developers. Longitudinal studies to track the evolution of guideline comprehension over time, assessing how guideline adherence influences code maintainability, and extending the analysis to style guides across various programming languages are also recommended to enhance both educational strategies and software development practices.
\revision{
We intend to investigate more extensive code and the impact of guidelines on gender and neurodiversity.
We aim at analyzing the \textit{Comparison to True} guideline further by interviewing more undergraduate students to better understand the results.
}

%% file: sigconf.bbl

\begin{thebibliography}{37}


\ifx \showCODEN    \undefined \def \showCODEN     #1{\unskip}     \fi
\ifx \showDOI      \undefined \def \showDOI       #1{#1}\fi
\ifx \showISBNx    \undefined \def \showISBNx     #1{\unskip}     \fi
\ifx \showISBNxiii \undefined \def \showISBNxiii  #1{\unskip}     \fi
\ifx \showISSN     \undefined \def \showISSN      #1{\unskip}     \fi
\ifx \showLCCN     \undefined \def \showLCCN      #1{\unskip}     \fi
\ifx \shownote     \undefined \def \shownote      #1{#1}          \fi
\ifx \showarticletitle \undefined \def \showarticletitle #1{#1}   \fi
\ifx \showURL      \undefined \def \showURL       {\relax}        \fi
\providecommand\bibfield[2]{#2}
\providecommand\bibinfo[2]{#2}
\providecommand\natexlab[1]{#1}
\providecommand\showeprint[2][]{arXiv:#2}

\bibitem[Allamanis et~al\mbox{.}(2014)]%
        {allamanis2014learning}
\bibfield{author}{\bibinfo{person}{Miltiadis Allamanis}, \bibinfo{person}{Earl~T Barr}, \bibinfo{person}{Christian Bird}, {and} \bibinfo{person}{Charles Sutton}.} \bibinfo{year}{2014}\natexlab{}.
\newblock \showarticletitle{Learning {N}atural {C}oding {C}onventions}. In \bibinfo{booktitle}{\emph{Proceedings of the International Symposium on Foundations of Software Engineering (FSE'14)}}. \bibinfo{pages}{281--293}.
\newblock


\bibitem[Basili et~al\mbox{.}(1994)]%
        {basili-1994}
\bibfield{author}{\bibinfo{person}{Victor Basili}, \bibinfo{person}{G. Caldiera}, {and} \bibinfo{person}{H. Rombach}.} \bibinfo{year}{1994}\natexlab{}.
\newblock \showarticletitle{The {G}oal {Q}uestion {M}etric {A}pproach}.
\newblock \bibinfo{journal}{\emph{Encyclopedia of {S}oftware {E}ngineering}} (\bibinfo{year}{1994}), \bibinfo{pages}{528--532}.
\newblock


\bibitem[Bauer et~al\mbox{.}(2019)]%
        {bauer2019indentation}
\bibfield{author}{\bibinfo{person}{Jennifer Bauer}, \bibinfo{person}{Janet Siegmund}, \bibinfo{person}{Norman Peitek}, \bibinfo{person}{Johannes~C. Hofmeister}, {and} \bibinfo{person}{Sven Apel}.} \bibinfo{year}{2019}\natexlab{}.
\newblock \showarticletitle{Indentation: {A}imply a {M}atter of {S}tyle or {S}upport for {P}rogram {C}omprehension?}. In \bibinfo{booktitle}{\emph{Proceedings of the International Conference on Program Comprehension}} \emph{(\bibinfo{series}{ICPC'19})}. IEEE, \bibinfo{pages}{154--164}.
\newblock


\bibitem[Binkley et~al\mbox{.}(2013)]%
        {binkley2013impact}
\bibfield{author}{\bibinfo{person}{Dave Binkley}, \bibinfo{person}{Marcia Davis}, \bibinfo{person}{Dawn Lawrie}, \bibinfo{person}{Jonathan Maletic}, \bibinfo{person}{Christopher Morrell}, {and} \bibinfo{person}{Bonita Sharif}.} \bibinfo{year}{2013}\natexlab{}.
\newblock \showarticletitle{The impact of {I}dentifier {S}tyle on {E}ffort and {C}omprehension}.
\newblock \bibinfo{journal}{\emph{Empirical Software Engineering}} \bibinfo{volume}{18}, \bibinfo{number}{2} (\bibinfo{year}{2013}), \bibinfo{pages}{219--276}.
\newblock


\bibitem[Box et~al\mbox{.}(2005)]%
        {BO05ST}
\bibfield{author}{\bibinfo{person}{George Box}, \bibinfo{person}{J.~Stuart Hunter}, {and} \bibinfo{person}{William~G. Hunter}.} \bibinfo{year}{2005}\natexlab{}.
\newblock \bibinfo{booktitle}{\emph{Statistics for {E}xperimenters}}.
\newblock \bibinfo{publisher}{Wiley-Interscience}.
\newblock


\bibitem[Buse and Weimer(2009)]%
        {buse2010learning}
\bibfield{author}{\bibinfo{person}{Raymond Buse} {and} \bibinfo{person}{Westley Weimer}.} \bibinfo{year}{2009}\natexlab{}.
\newblock \showarticletitle{Learning a Metric for Code Readability}. In \bibinfo{booktitle}{\emph{Proceedings of the International Symposium on Software Testing and Analysis}}. \bibinfo{pages}{465--475}.
\newblock


\bibitem[Buse and Weimer(2008)]%
        {buse2008metric}
\bibfield{author}{\bibinfo{person}{Raymond P.~L. Buse} {and} \bibinfo{person}{Westley~R. Weimer}.} \bibinfo{year}{2008}\natexlab{}.
\newblock \showarticletitle{A Metric for Software Readability}. In \bibinfo{booktitle}{\emph{Proceedings of the 2008 International Symposium on Software Testing and Analysis (ISSTA'08)}}. ACM Press, \bibinfo{pages}{121--130}.
\newblock


\bibitem[Busjahn et~al\mbox{.}(2015)]%
        {busjahn2015eye}
\bibfield{author}{\bibinfo{person}{Teresa Busjahn}, \bibinfo{person}{Carsten Schulte}, \bibinfo{person}{Sascha Tamm}, {and} \bibinfo{person}{Roman Bednarik}.} \bibinfo{year}{2015}\natexlab{}.
\newblock \showarticletitle{Eye {M}ovements in {P}rogramming {E}ducation {II}: Analyzing the {N}ovice's {G}aze}. In \bibinfo{booktitle}{\emph{Proceedings of the Conference on Computing Education (ICER'15)}}.
\newblock


\bibitem[Crosby et~al\mbox{.}(2002)]%
        {CR02roles}
\bibfield{author}{\bibinfo{person}{Martha Crosby}, \bibinfo{person}{Jean Scholtz}, {and} \bibinfo{person}{Susan Wiedenbeck}.} \bibinfo{year}{2002}\natexlab{}.
\newblock \showarticletitle{The {R}oles {B}eacons {P}lay in {C}omprehension for {N}ovice and {E}xpert {P}rogrammers.}. In \bibinfo{booktitle}{\emph{Workshop of the Psychology of Programming Interest Group}} \emph{(\bibinfo{series}{PPIG'02})}. \bibinfo{pages}{5}.
\newblock


\bibitem[da~Costa and Gheyi(2023)]%
        {CostaGheyi2023}
\bibfield{author}{\bibinfo{person}{José Aldo~Silva da Costa} {and} \bibinfo{person}{Rohit Gheyi}.} \bibinfo{year}{2023}\natexlab{}.
\newblock \showarticletitle{Evaluating the Code Comprehension of Novices with Eye Tracking}. In \bibinfo{booktitle}{\emph{Concurso de Teses e Dissertações em Engenharia de Software (CTD-ES)}}.
\newblock


\bibitem[da~Costa et~al\mbox{.}(2023)]%
        {da2023seeing}
\bibfield{author}{\bibinfo{person}{Jos{\'e} Aldo~Silva da Costa}, \bibinfo{person}{Rohit Gheyi}, \bibinfo{person}{Fernando Castor}, \bibinfo{person}{Pablo Roberto~Fernandes de Oliveira}, \bibinfo{person}{M{\'a}rcio Ribeiro}, {and} \bibinfo{person}{Baldoino Fonseca}.} \bibinfo{year}{2023}\natexlab{}.
\newblock \showarticletitle{Seeing {C}onfusion through a {N}ew {L}ens: on the {I}mpact of {A}toms of {C}onfusion on {N}ovices’ {C}ode {C}omprehension}.
\newblock \bibinfo{journal}{\emph{Empirical Software Engineering}} \bibinfo{volume}{28}, \bibinfo{number}{4} (\bibinfo{year}{2023}), \bibinfo{pages}{81}.
\newblock


\bibitem[da~Costa et~al\mbox{.}(2021)]%
        {costa2021evaluating}
\bibfield{author}{\bibinfo{person}{Jos{\'e} Aldo~Silva da Costa}, \bibinfo{person}{Rohit Gheyi}, \bibinfo{person}{M{\'a}rcio Ribeiro}, \bibinfo{person}{Sven Apel}, \bibinfo{person}{Vander Alves}, \bibinfo{person}{Baldoino Fonseca}, \bibinfo{person}{Fl{\'a}vio Medeiros}, {and} \bibinfo{person}{Alessandro Garcia}.} \bibinfo{year}{2021}\natexlab{}.
\newblock \showarticletitle{Evaluating {R}efactorings for {D}isciplining \#ifdef {A}nnotations: An {E}ye {T}racking {S}tudy with {N}ovices}.
\newblock \bibinfo{journal}{\emph{Empirical Software Engineering}} \bibinfo{volume}{26}, \bibinfo{number}{5} (\bibinfo{year}{2021}), \bibinfo{pages}{1--35}.
\newblock


\bibitem[Daka et~al\mbox{.}(2015)]%
        {daka2015modeling}
\bibfield{author}{\bibinfo{person}{Ermira Daka}, \bibinfo{person}{Jos{\'e} Campos}, \bibinfo{person}{Gordon Fraser}, \bibinfo{person}{Jonathan Dorn}, {and} \bibinfo{person}{Westley Weimer}.} \bibinfo{year}{2015}\natexlab{}.
\newblock \showarticletitle{Modeling readability to improve unit tests}. In \bibinfo{booktitle}{\emph{Proceedings of the Foundations of Software Engineering}}. \bibinfo{pages}{107--118}.
\newblock


\bibitem[Dasgupta and Hooshangi(2017)]%
        {dasgupta2017code}
\bibfield{author}{\bibinfo{person}{Subhasish Dasgupta} {and} \bibinfo{person}{Sara Hooshangi}.} \bibinfo{year}{2017}\natexlab{}.
\newblock \showarticletitle{Code {Q}uality: {E}xamining the {E}fficacy of {A}utomated {T}ools}. In \bibinfo{booktitle}{\emph{Americas Conference on Information Systems (AMCIS'17)}}.
\newblock


\bibitem[de~Almeida et~al\mbox{.}(2003)]%
        {de2003best}
\bibfield{author}{\bibinfo{person}{Jorgy~Rady de Almeida}, \bibinfo{person}{Jo\~{a}o~Batista Camargo}, \bibinfo{person}{Bruno~Abrantes Basseto}, {and} \bibinfo{person}{S{\'e}rgio~Miranda Paz}.} \bibinfo{year}{2003}\natexlab{}.
\newblock \showarticletitle{Best {P}ractices in {C}ode {I}nspection for {S}afety-critical {S}oftware}.
\newblock \bibinfo{journal}{\emph{IEEE Software}} \bibinfo{volume}{20}, \bibinfo{number}{3} (\bibinfo{year}{2003}), \bibinfo{pages}{56--63}.
\newblock


\bibitem[de~Oliveira et~al\mbox{.}(2020)]%
        {oliveira2020atoms}
\bibfield{author}{\bibinfo{person}{Benedito de Oliveira}, \bibinfo{person}{M{\'a}rcio Ribeiro}, \bibinfo{person}{Jos{\'e} Aldo~Silva da Costa}, \bibinfo{person}{Rohit Gheyi}, \bibinfo{person}{Guilherme Amaral}, \bibinfo{person}{Rafael de Mello}, \bibinfo{person}{Anderson Oliveira}, \bibinfo{person}{Alessandro Garcia}, \bibinfo{person}{Rodrigo Bonif{\'a}cio}, {and} \bibinfo{person}{Baldoino Fonseca}.} \bibinfo{year}{2020}\natexlab{}.
\newblock \showarticletitle{Atoms of {C}onfusion: {T}he {E}yes {D}o {N}ot {L}ie}. In \bibinfo{booktitle}{\emph{Proceedings of the Brazilian Symposium on Software Engineering}} \emph{(\bibinfo{series}{SBES'20})}. \bibinfo{pages}{243--252}.
\newblock


\bibitem[dos Santos and Gerosa(2018)]%
        {dos2018impacts}
\bibfield{author}{\bibinfo{person}{Rodrigo~Magalh{\~a}es dos Santos} {and} \bibinfo{person}{Marco~Aur{\'e}lio Gerosa}.} \bibinfo{year}{2018}\natexlab{}.
\newblock \showarticletitle{Impacts of {C}oding {P}ractices on {R}eadability}. In \bibinfo{booktitle}{\emph{Proceedings of the International Conference on Program Comprehension (ICPC'18)}}. \bibinfo{pages}{277--285}.
\newblock


\bibitem[Fakhoury et~al\mbox{.}(2020)]%
        {Fakhoury2020}
\bibfield{author}{\bibinfo{person}{Sarah Fakhoury}, \bibinfo{person}{Devjeet Roy}, \bibinfo{person}{Yuzhan Ma}, \bibinfo{person}{Venera Arnaoudova}, {and} \bibinfo{person}{Olusola Adesope}.} \bibinfo{year}{2020}\natexlab{}.
\newblock \showarticletitle{Measuring the impact of lexical and structural inconsistencies on developers' cognitive load during bug localization}.
\newblock \bibinfo{journal}{\emph{Empirical Software Engineering}}  \bibinfo{volume}{25} (\bibinfo{year}{2020}), \bibinfo{pages}{2140--2178}.
\newblock


\bibitem[Falessi et~al\mbox{.}(2018)]%
        {DBLP:journals/ese/FalessiJWTMJO18}
\bibfield{author}{\bibinfo{person}{Davide Falessi}, \bibinfo{person}{Natalia Juristo}, \bibinfo{person}{Claes Wohlin}, \bibinfo{person}{Burak Turhan}, \bibinfo{person}{J{\"{u}}rgen M{\"{u}}nch}, \bibinfo{person}{Andreas Jedlitschka}, {and} \bibinfo{person}{Markku Oivo}.} \bibinfo{year}{2018}\natexlab{}.
\newblock \showarticletitle{Empirical software engineering experts on the use of students and professionals in experiments}.
\newblock \bibinfo{journal}{\emph{Empirical Software Engineering}} \bibinfo{volume}{23}, \bibinfo{number}{1} (\bibinfo{year}{2018}), \bibinfo{pages}{452--489}.
\newblock


\bibitem[{Google}(2024)]%
        {google-styleguide}
\bibfield{author}{\bibinfo{person}{{Google}}.} \bibinfo{year}{2024}\natexlab{}.
\newblock \bibinfo{booktitle}{\emph{{Google Python Style Guide}}}.
\newblock
\urldef\tempurl%
\url{https://google.github.io/styleguide/pyguide.html}
\showURL{%
\tempurl}


\bibitem[Lawrie et~al\mbox{.}(2006)]%
        {lawrie2006name}
\bibfield{author}{\bibinfo{person}{D. Lawrie}, \bibinfo{person}{C. Morrell}, \bibinfo{person}{H. Feild}, {and} \bibinfo{person}{D. Binkley}.} \bibinfo{year}{2006}\natexlab{}.
\newblock \showarticletitle{What’s in a name? {A} study of identifiers}. In \bibinfo{booktitle}{\emph{14th IEEE International Conference on Program Comprehension (ICPC'06)}}. IEEE, \bibinfo{pages}{3--12}.
\newblock


\bibitem[Lawrie et~al\mbox{.}(2007)]%
        {lawrie2007effective}
\bibfield{author}{\bibinfo{person}{Dawn Lawrie}, \bibinfo{person}{Christopher Morrell}, \bibinfo{person}{Henry Feild}, {and} \bibinfo{person}{David Binkley}.} \bibinfo{year}{2007}\natexlab{}.
\newblock \showarticletitle{Effective Identifier Names for Comprehension and Memory}.
\newblock \bibinfo{journal}{\emph{Innovations in Systems and Software Engineering}}  \bibinfo{volume}{3} (\bibinfo{year}{2007}), \bibinfo{pages}{303--318}.
\newblock


\bibitem[{Microsoft}(2024)]%
        {microsoft-python-formatting}
\bibfield{author}{\bibinfo{person}{{Microsoft}}.} \bibinfo{year}{2024}\natexlab{}.
\newblock \bibinfo{booktitle}{\emph{{Formatting Python Code}}}.
\newblock
\urldef\tempurl%
\url{https://learn.microsoft.com/en-us/visualstudio/python/formatting-python-code?view=vs-2022}
\showURL{%
\tempurl}


\bibitem[Nystr{\"o}m and Holmqvist(2010)]%
        {nystrom2010adaptive}
\bibfield{author}{\bibinfo{person}{Marcus Nystr{\"o}m} {and} \bibinfo{person}{Kenneth Holmqvist}.} \bibinfo{year}{2010}\natexlab{}.
\newblock \showarticletitle{An adaptive algorithm for fixation, saccade, and glissade detection in eyetracking data}.
\newblock \bibinfo{journal}{\emph{Behavior {r}esearch {m}ethods}} \bibinfo{volume}{42}, \bibinfo{number}{1} (\bibinfo{year}{2010}), \bibinfo{pages}{188--204}.
\newblock


\bibitem[Oliveira et~al\mbox{.}(2020)]%
        {Oliveira2020evaluating}
\bibfield{author}{\bibinfo{person}{Delano Oliveira}, \bibinfo{person}{Reydne Bruno}, \bibinfo{person}{Fernanda Madeiral}, {and} \bibinfo{person}{Fernando Castor}.} \bibinfo{year}{2020}\natexlab{}.
\newblock \showarticletitle{Evaluating {C}ode {R}eadability and {L}egibility: An {E}xamination of {H}uman-centric {S}tudies}. In \bibinfo{booktitle}{\emph{Proceedings of the International Conference on Software Maintenance and Evolution}} \emph{(\bibinfo{series}{ICSME'20})}. \bibinfo{pages}{348--359}.
\newblock


\bibitem[Posnett et~al\mbox{.}(2011)]%
        {posnett2011simpler}
\bibfield{author}{\bibinfo{person}{D. Posnett}, \bibinfo{person}{A. Hindle}, {and} \bibinfo{person}{P. Devanbu}.} \bibinfo{year}{2011}\natexlab{}.
\newblock \showarticletitle{A Simpler Model of Software Readability}. In \bibinfo{booktitle}{\emph{Proceedings of the 8th Working Conference on Mining Software Repositories (MSR'11)}}. ACM Press, \bibinfo{pages}{73--82}.
\newblock


\bibitem[Rayner(1998)]%
        {rayner1998eye}
\bibfield{author}{\bibinfo{person}{Keith Rayner}.} \bibinfo{year}{1998}\natexlab{}.
\newblock \showarticletitle{Eye {M}ovements in {R}eading and {I}nformation {P}rocessing: 20 {Y}ears of {R}esearch}.
\newblock \bibinfo{journal}{\emph{Psychological {B}ulletin}} \bibinfo{volume}{124}, \bibinfo{number}{3} (\bibinfo{year}{1998}), \bibinfo{pages}{372}.
\newblock


\bibitem[Roman(2020)]%
        {sharafi2020practical}
\bibfield{author}{\bibinfo{person}{Sharafi Zohreh; Bonita Sharif; Yann-Ga\"{e}l Gu\'{e}h\'{e}neuc; Andrew Begel;~Bednarik Roman}.} \bibinfo{year}{2020}\natexlab{}.
\newblock \showarticletitle{A practical guide on conducting eye tracking studies in software engineering}.
\newblock \bibinfo{journal}{\emph{Empirical Software Engineering}}  \bibinfo{volume}{25} (\bibinfo{year}{2020}), \bibinfo{pages}{3128--3174}.
\newblock


\bibitem[Salman et~al\mbox{.}(2015)]%
        {DBLP:conf/icse/SalmanMJ15}
\bibfield{author}{\bibinfo{person}{Iflaah Salman}, \bibinfo{person}{Ayse~Tosun Misirli}, {and} \bibinfo{person}{Natalia~Juristo Juzgado}.} \bibinfo{year}{2015}\natexlab{}.
\newblock \showarticletitle{Are Students Representatives of Professionals in Software Engineering Experiments?}. In \bibinfo{booktitle}{\emph{37th {IEEE/ACM} International Conference on Software Engineering, {ICSE} 2015, Florence, Italy, May 16-24, 2015, Volume 1}}. \bibinfo{publisher}{{IEEE} Computer Society}, \bibinfo{pages}{666--676}.
\newblock


\bibitem[Salvucci and Goldberg(2000)]%
        {salvucci2000identifying}
\bibfield{author}{\bibinfo{person}{Dario Salvucci} {and} \bibinfo{person}{Joseph Goldberg}.} \bibinfo{year}{2000}\natexlab{}.
\newblock \showarticletitle{Identifying {F}ixations and {S}accades in {E}ye-tracking {P}rotocols}. In \bibinfo{booktitle}{\emph{Proceedings of the Symposium on Eye Tracking Research \& Applications}} \emph{(\bibinfo{series}{ETRA'00})}. \bibinfo{pages}{71--78}.
\newblock


\bibitem[Santos(2021)]%
        {santos2021estudo}
\bibfield{author}{\bibinfo{person}{Reydne Bruno~dos Santos}.} \bibinfo{year}{2021}\natexlab{}.
\newblock \emph{\bibinfo{title}{Um {E}studo sobre {D}efini{\c{c}}{\~a}o e {A}valia{\c{c}}{\~a}o da {R}eadability e {L}egibility do {C}{\'o}digo {F}onte}}.
\newblock \bibinfo{thesistype}{Master's\ thesis}. \bibinfo{school}{Universidade Federal de Pernambuco}.
\newblock


\bibitem[Schankin et~al\mbox{.}(2018)]%
        {schankin2018descriptive}
\bibfield{author}{\bibinfo{person}{Andrea Schankin}, \bibinfo{person}{Annika Berger}, \bibinfo{person}{Daniel~V. Holt}, \bibinfo{person}{Johannes~C. Hofmeister}, \bibinfo{person}{Till Riedel}, {and} \bibinfo{person}{Michael Beigl}.} \bibinfo{year}{2018}\natexlab{}.
\newblock \showarticletitle{Descriptive {C}ompound {I}dentifier {N}ames {I}mprove {S}ource {C}ode {C}omprehension}. In \bibinfo{booktitle}{\emph{Proceedings of the International Conference on Program Comprehension (ICPC'18)}}. \bibinfo{pages}{31--40}.
\newblock


\bibitem[Sharafi et~al\mbox{.}(2021)]%
        {Sharafi2021}
\bibfield{author}{\bibinfo{person}{Zohreh Sharafi}, \bibinfo{person}{Yu Huang}, \bibinfo{person}{Kevin Leach}, {and} \bibinfo{person}{Westley Weimer}.} \bibinfo{year}{2021}\natexlab{}.
\newblock \showarticletitle{Toward an Objective Measure of Developers’ Cognitive Activities}.
\newblock \bibinfo{journal}{\emph{ACM Transactions on Software Engineering and Methodology}} \bibinfo{volume}{30}, \bibinfo{number}{3} (\bibinfo{year}{2021}), \bibinfo{pages}{1--40}.
\newblock


\bibitem[Sharafi et~al\mbox{.}(2012)]%
        {sharafi2012women}
\bibfield{author}{\bibinfo{person}{Zohreh Sharafi}, \bibinfo{person}{Z{\'e}phyrin Soh}, \bibinfo{person}{Yann-Ga{\"e}l Gu{\'e}h{\'e}neuc}, {and} \bibinfo{person}{Giuliano Antoniol}.} \bibinfo{year}{2012}\natexlab{}.
\newblock \showarticletitle{Women and {M}en—{D}ifferent but {E}qual: On the {I}mpact of {I}dentifier {S}tyle on {S}ource {C}ode {R}eading}. In \bibinfo{booktitle}{\emph{Proceedings of the International Conference on Program Comprehension}} \emph{(\bibinfo{series}{ICPC'12})}. IEEE, \bibinfo{pages}{27--36}.
\newblock


\bibitem[Sharif and Maletic(2010)]%
        {sharif2010aneye}
\bibfield{author}{\bibinfo{person}{Bonita Sharif} {and} \bibinfo{person}{Jonathan Maletic}.} \bibinfo{year}{2010}\natexlab{}.
\newblock \showarticletitle{An {E}ye {T}racking {S}tudy on {C}amelcase and {U}nder\_score {I}dentifier {S}tyles}. In \bibinfo{booktitle}{\emph{Proceedings of the International Conference on Program Comprehension}} \emph{(\bibinfo{series}{ICPC'10})}. IEEE, \bibinfo{pages}{196--205}.
\newblock


\bibitem[Stefik and Siebert(2013)]%
        {stefik2013empirical}
\bibfield{author}{\bibinfo{person}{Andreas Stefik} {and} \bibinfo{person}{Susanna Siebert}.} \bibinfo{year}{2013}\natexlab{}.
\newblock \showarticletitle{An {E}mpirical {I}nvestigation into {P}rogramming {L}anguage {S}yntax}.
\newblock \bibinfo{journal}{\emph{ACM Transactions on Computing Education (TOCE'13)}} \bibinfo{volume}{13}, \bibinfo{number}{4} (\bibinfo{year}{2013}), \bibinfo{pages}{1--40}.
\newblock


\bibitem[Van~Rossum et~al\mbox{.}(2001)]%
        {van2001pep}
\bibfield{author}{\bibinfo{person}{Guido Van~Rossum}, \bibinfo{person}{Barry Warsaw}, {and} \bibinfo{person}{Nick Coghlan}.} \bibinfo{year}{2001}\natexlab{}.
\newblock \showarticletitle{{PEP}8--{S}tyle{G}uide for {P}ython {C}ode}.
\newblock \bibinfo{journal}{\emph{Python.org}}  \bibinfo{volume}{1565} (\bibinfo{year}{2001}), \bibinfo{pages}{28}.
\newblock


\end{thebibliography}
